\documentclass[twocolumn]{aastex631}

\shorttitle{chondrule destrucion in shock}
\shortauthors{Matsumoto, Kurosawa, Arakawa}

\newcommand{\erf}[1]{{\rm erf}{{#1}}}
\newcommand{\We}{\mathcal We}
\newcommand{\Oh}{\mathcal Oh}

\begin{document}

\title{Chondrule Destruction via Dust Collisions in Shock Waves}

\correspondingauthor{Yuji Matsumoto}
\email{yuji.matsumoto@nao.ac.jp}

\author[0000-0002-2383-1216]{Yuji Matsumoto}
\affiliation{National Astronomical Observatory of Japan, 2-21-1, Osawa, Mitaka, 181-8588 Tokyo, Japan}

\author[0000-0003-4965-4585]{Kosuke Kurosawa}
\affiliation{Department of Human Environmental Science, Graduate school of Human Development and Environment, Kobe University, 3-11, Tsurukabuto, Nada-ku, Kobe, Hyogo 657-8501, Japan}
\affiliation{Planetary Exploration Research Center, Chiba Institute of Technology, 2-17-1, Narashino, Tsudanuma, Chiba 275-0016, Japan}

\author[0000-0003-0947-9962]{Sota Arakawa}
\affiliation{Japan Agency for Marine-Earth Science and Technology, 3173-25, Showa-machi, Kanazawa-ku, Yokohama, Kanagawa 236-0001, Japan}

\begin{abstract}
    A leading candidate for the heating source of chondrules and igneous rims is shock waves.
    This mechanism generates high relative velocities between chondrules and dust particles.
    We have investigated the possibility of the chondrule destruction in collisions with dust particles behind a shock wave using a semianalytical treatment.
    We find that the chondrules are destroyed during melting in collisions.
    We derive the conditions for the destruction of chondrules and show that the typical size of the observed chondrules satisfies the condition.
    We suggest that the chondrule formation and rim accretion are different events if they are heated by shock waves.
\end{abstract}

	
\section{Introduction} \label{sect:intro}

Chondrules are spherical-shaped igneous grains in chondrites.
They were once wholly or partially molten by flash heating events in the solar nebula \citep[e.g.,][]{Hewins+2005}.
Chondrules are the most abundant ingredients of most chondrites, suggesting that flash heating events in the solar nebula were frequent.
The heating processes of chondrules are still under debate.
Several heating mechanisms have been proposed so far, including shock waves \citep[e.g.,][]{Hood&Horanyi1991,Miura&Nakamoto2005}, planetesimal collisions \citep[e.g.,][]{Asphaug+2011,Johnson+2015,Wakita+2017}, and lightning \citep[e.g., ][]{Horanyi+1995, Kaneko+2023}.

Rims are the surrounding structures of chondrules.
There are two major categories pertaining to rims: fine-grained rims and igneous rims.
Fine-grained rims are composed of matrix-like dust grains \citep[e.g.,][]{Ashworth1977, Huss+2005}.
The origin of the fine-grained rims is mainly interpreted as the accretion of dust particles onto the surfaces of chondrules \citep[e.g.,][]{Metzler+1992,Matsumoto+2019, Kaneko+2022}.
Igneous rims are composed of coarse-grained dust particles and show evidence of a high degree of melting \citep[e.g.,][]{Rubin1984,Krot&Wasson1995}.
The dust particles in the igneous rim components underwent flash heating events similar to those that formed the chondrule.
Igneous rims would be formed by melting of accretionary rims \citep[e.g.,][]{Rubin1984,Rubin2010} and the accretion of droplets \citep{Kring1991,Jacquet+2013, Matsumoto+2021,Matsumoto&Arakawa2023}.

The presence of the rims suggests that chondrules and dust particles coexisted in the solar nebula. 
Chondrules accrete dust particles during or after heating events. 
The accretion conditions of the chondrule-dust collisions depend on the impact velocity and their physical states.
In particular, chondrules are expected to be destroyed if the impact velocity is too high.

Previous studies have pointed out the possibility of chondrule destruction in shock heating events.
In this flash heating process, chondrules are decelerated by the gas drag force in the post-shock regions \citep[e.g.,][]{Hood&Horanyi1991,Iida+2001}.
Dust particles of different sizes have relative velocities due to the size dependence of the stopping time.
\cite{Nakamoto&Miura2004_LPSC} evaluated the collisional destruction rate of chondrules adopting the destruction criterion based on the impact fragmentation experiment \citep[][]{Takagi_Y+1984}.
They showed that chondrules larger than a critical radius are destroyed.
\cite{Jacquet&Thompson_C2014} pointed out that impact fragmentation debris promotes the destruction of chondrules. 
\cite{Ciesla2006} considered collisions between molten chondrules.
Molten chondrules have different criteria for adhesion and destruction from the chondrules after crystallization. 
\cite{Ciesla2006} estimated the impact velocity for the destruction of molten chondrules with viscosities of 100 poise (P) and showed that the molten chondrules are destroyed when the impact velocity is higher than $\sim100~\mbox{m~s}^{-1}$.
They also showed that chondrules can avoid destructive collisions during melting because only a small number of chondrules collide during melting.
Here we have revisited the possibility of the collisional disruption of chondrules behind a shock wave.
We propose a new scenario for the collisional disruption of chondrules (or chondrule precursors): the destruction of molten chondrules by collisions with $\sim10~\mu$m dust particles.
Our results suggest that the chondrule-forming shock events occur in $\sim10~\mu$m dust-free environments, while the rim-accreting shock events occur in the presence of $\sim10~\mu$m dust particles.
The plan of our paper is as follows.
We describe our model in Section \ref{sect:model}.
We show the typical evolution of chondrules and the parameter dependence in Section \ref{sect:results}.
Our conclusion and discussion is given in Section \ref{sect:conclusion}, where we discuss the conditions of the chondrule-forming shock and the rim-accreting shock.

\section{Model} \label{sect:model}

Chondrules (precursors) and dust particles are decelerated and heated behind the shock front. 
A chondrule collides with dust particles according to the number density of the dust, the collisional cross section, and their relative velocities.
We adopt the one-dimensional normal shock model used in \cite{Matsumoto&Arakawa2023} to consider the destruction of chondrules in collisions.

\subsection{Gas Structure}

A simple gas structure is assumed according to previous studies \citep[e.g.,][]{Nakamoto&Miura2004_LPSC,Ciesla2006,Jacquet&Thompson_C2014,Arakawa&Nakamoto2019,Matsumoto&Arakawa2023}. 
Here, we briefly describe the key equations with the numbers used in the model. 
The details can be found in the original studies. 
The gas velocity with respect to the shock front, $v_g$, is given by 
\begin{eqnarray}
    v_g = 
    \left\{
        \begin{array}{l}
            v_0 \hfill(x<0),\\
            v_0 + (v_{\rm post} -v_0) \exp{(-x/L)}\quad \hfill (x\geq0),
        \end{array}
      \right.
      \label{eq:v_g}
\end{eqnarray}
where $v_0$ is the gas velocity with respect to the shock front in the preshock region, namely, shock velocity, $v_{\rm post}$ is the postshock gas velocity with respect to the shock front, $L=10^3$~km \citep{Miura&Nakamoto2005} is the spatial scale of the shock, and $x$ is the distance from the shock front.
The postshock gas velocity is given by the Rankine--Hugoniot relations, $v_{\rm post} = [(\gamma-1)/(\gamma+1)]v_0$, where the ratio of specific heat is $\gamma=1.4$ for the nebula gas composed of molecular hydrogen.
The gas temperature is
\begin{eqnarray}
    T_g = 
    \left\{
        \begin{array}{l}
            T_0 \hfill(x<0),\\
            T_0 + (T_{\rm post} -T_0) \exp{(-x/L)}\quad \hfill (x\geq0),
        \end{array}
      \right.
\end{eqnarray}
where the preshock temperature is $T_0=500$~K and the postshock gas temperature is $T_{\rm post}=2000$~K \citep{Miura+2002}.
According to the Rankine--Hugoniot and isobaric relations, the gas number density can be expressed as
\begin{eqnarray}
    n_g = n_0 \frac{T_0}{T_g} \frac{ 4 s_0^2 - (\gamma-1)}{\gamma+1},
    \label{eq:n_g}
\end{eqnarray}
where $n_0=10^{14}~\mbox{cm}^{-3}$ is the preshock gas number density, $s_0=v_0/(2k_{\rm B} T_{0}/m_g )^{1/2}$, $k_{\rm B}$ is the Boltzmann constant, and $m_g=3.34\times 10^{-24}$~g is the mass of H$_2$.

\subsection{Dynamical and Temperature Evolution of Chondrules and Dust}

Chondrules and dust particles are decelerated once and then accelerated by interaction with the gas behind the shock front.
Their velocities with respect to the shock front are given by
\begin{eqnarray}
    m \frac{d v}{d x} = -\frac{ C_D }{2} \pi a^2 m_g n_g \frac{|v-v_g|}{v} (v-v_g),
\end{eqnarray}
where $v$, $m$, and $a$ are the velocity, mass, and radius of a chondrule (ch) or dust particle ($d$), respectively.
The radii of the chondrules are our parameters and $a_d=20~\mu$m.
The density of the chondrules and dust particles are $\rho=3.3~\mbox{g~cm}^{-3}$ \citep[e.g.,][]{Friedrich+2015}.
The drag coefficient, $C_D$, is given by \citep{Hood&Horanyi1991}
\begin{eqnarray}
    C_D &\simeq&
    \frac{2\sqrt{\pi}}{3s} + \frac{2s^2+1}{\sqrt{\pi} s^3} \exp{(-s^2)} 
    + \frac{4s^4+2s^2-1}{2s^4} \erf{(s)}.
    \nonumber\\
\end{eqnarray}
where $s = v /(2k_{\rm B} T_{\rm g}/m_{\rm g} )^{1/2}$.
The stopping length of a particle becomes 
\begin{eqnarray}
    l_{\rm stop}
    &\equiv& \left( \frac{1}{v} \frac{dv}{dx} \right)^{-1}
    \nonumber\\
    &\simeq &
    1.3 \times 10^2 ~\mbox{km}
    \left( \frac{ a }{ 100~\mbox{$\mu$m}} \right)
    \left( \frac{n_0}{10^{14}\mbox{~cm}^{-3}} \right)^{-1}
    \nonumber\\ && \times
    \left( \frac{v_0}{ 10\mbox{~km~s}^{-1}} \right)^{-2}
    \left( \frac{v}{v-v_g } \right)^2.
\end{eqnarray}

The temperature evolution of a particle is given by the equation of energy, 
\begin{eqnarray}
    m c_{\rm heat} \frac{{\rm d} T}{{\rm d} x} =
    \frac{4\pi a^2}{v} (\Gamma - \Lambda_{\rm rad} - \Lambda_{\rm evap} ),
    \label{eq:energy}
\end{eqnarray}
where $c_{\rm heat}=1.42\times 10^7~\mbox{erg~g$^{-1}$~K$^{-1}$}$ is the specific heat, which is taken from the value of forsterite\footnote{We note that actual chondrules are not composed entirely of forsterite \citep[e.g.,][]{Scott2007}.} \citep[][based on NIST WebBook]{Miura+2002}, $\Gamma$ is the heating rate via the energy transfer from the gas, $\Lambda_{\rm rad}$ is radiative cooling, and $\Lambda_{\rm evap}$ is latent heat cooling by evaporation per unit area, respectively.
The detailed expressions of $\Gamma$, $\Lambda_{\rm rad}$, and $\Lambda_{\rm evap}$ are described in Appendix \ref{sect:model_GLL}.
This equation indicates that smaller dust particles heat up more quickly.

Dust particles melt when their temperature reaches the melting temperature, $T_{\rm melt}=2171~$K \citep[][]{Miura+2002}.
We consider a thermal buffering by the latent heat from solid to liquid pertaining to both chondrules and dust particles with the mass fraction of solid, $S$, 
\begin{eqnarray}
    m L_{\rm melt} \frac{{\rm d} S}{{\rm d} x} =
    \frac{4\pi a^2}{v} (\Gamma - \Lambda_{\rm rad} - \Lambda_{\rm evap} ),
\end{eqnarray}
where $L_{\rm melt}=4.47\times 10^9~\mbox{erg~g}^{-1}$ is the latent heat of melting \citep[][]{Miura+2002}.
During the phase transition ($0<S<1$), we assume that the temperature of a particle remains constant. 
We assume that the crystallization temperature of droplets is $T_{\rm c}=1000$~K due to supercooling
\footnote{
According to experimental studies \citep[e.g.,][]{Nagashima_Ken+2008}, we consider that molten chondrules (precursors) turn into supercooled droplets.
The supercooled droplets crystallize at the crystallization temperature, which we take from the glass transition temperature \citep[$\approx900$--$1000$~K, e.g.,][]{Villeneuve+2015}.
The actual crystallization temperature depends on the composition and the cooling rate.
The glass transition temperature would be a lower limit of the crystallization temperature \citep[see also ][]{Arakawa&Nakamoto2016a}.
The crystallization temperature does not significantly affect the results.
}. 

We consider the size evolution of dust particles due to boiling and evaporation. 
The details are described in Appendix \ref{sect:ad_ev}.
The mass density of dust particles decreases as the size of a dust particle decreases.
Instead, we calculate the number density of dust particles. 
The number density of dust particles ($n_d$) is calculated from the dust-to-gas ratio in the preshock region ($\chi$) and the continuity equation in the steady state.

\subsection{Chondrule destruction}

\begin{figure}[ht!]
    \plotone{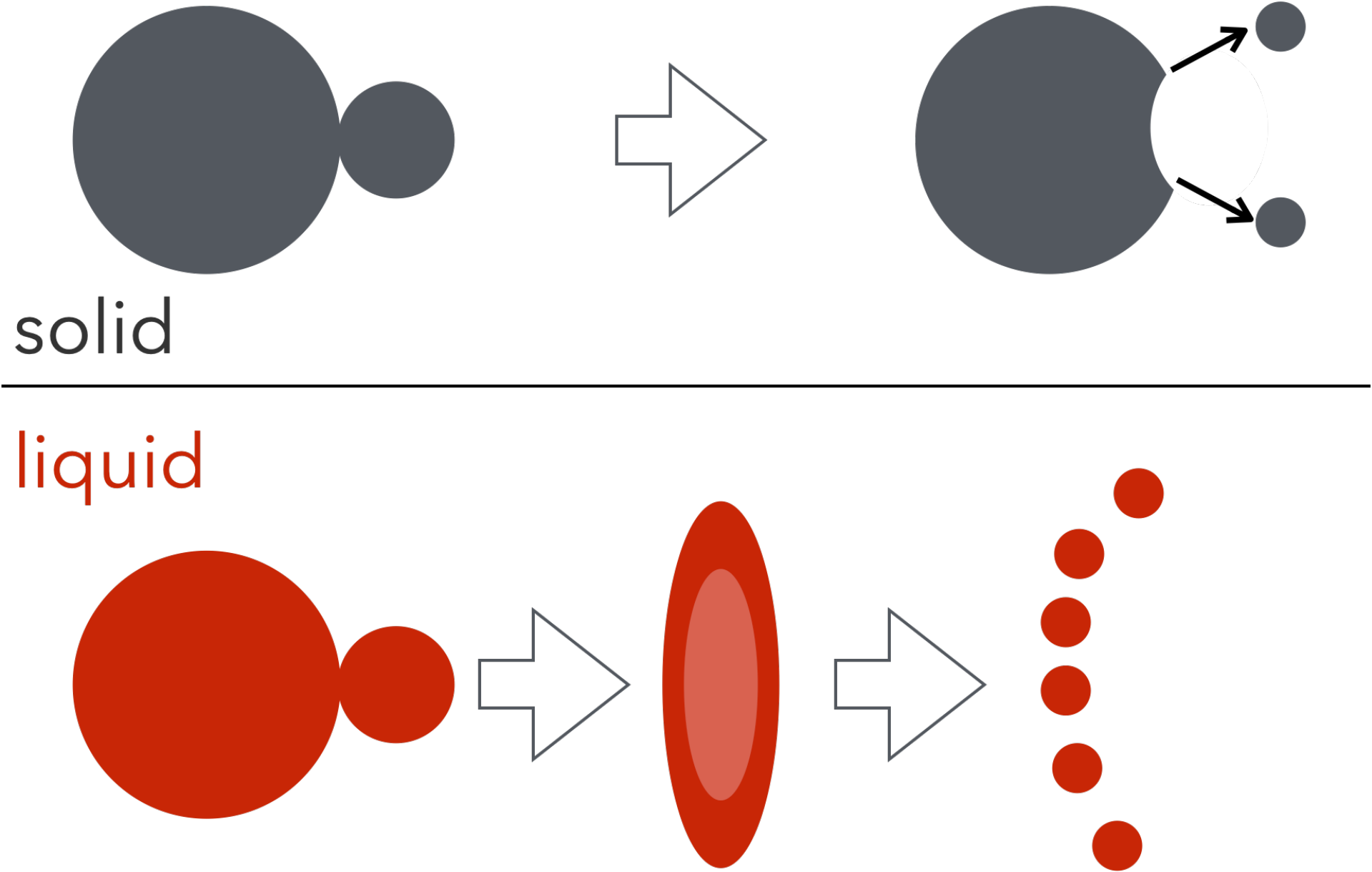}
    \caption{
        Schematics of the collision outcome models for the chondrules in this study.
        A solid chondrule is excavated by the collisions of dust particles \citep[e.g.,][]{Suzuki_A+2012}. 
        The droplet chondrule is destroyed by the collisions of dust particles as follows: 
        The droplet chondrule deforms, once becomes a lamella, and is broken into secondary droplets \citep[e.g.,][]{Pan_KL+2009}.
        \label{fig:schematic}}
\end{figure}

The destruction of chondrules in collisions is modeled based on impact excavation using experimental results on the crater scaling law, 
\begin{eqnarray}
    \pi_V &=& 0.11 \pi_3^{-0.71} \pi_4^{0.23},
\end{eqnarray}
where $\pi_V$ is the dimensionless volume, $\pi_3$ is the ratio of the strength to dynamic ram pressure, and $\pi_4$ is the density ratio \citep[e.g.,][]{Holsapple&Schmidt1982, Holsapple1993, Suzuki_A+2012},
\begin{eqnarray}
    \pi_V = \frac{\rho_t V}{m_p}, ~
    \pi_3 = \frac{Y}{\rho_p v_{\rm imp}^2}, ~
    \pi_4 = \frac{\rho_t}{\rho_p}.
\end{eqnarray}
The quantities with $t$ and $p$ are those of the target and projectile, $V$ is the crater volume, $Y$ is the target strength, and $v_{imp}$ is the impact velocity.
The strength of the solid chondrule is $Y_{\rm sol}=10~\mbox{MPa}$, according to the crushing strength of the chondrule from Allende and Saratov chondrites \citep[][]{Wada+2018}.
The mass-loss in a single impact excavation can be expressed as 
\begin{eqnarray}
    m_{\rm loss} &=& \pi_V m_{d} 
    \nonumber\\
    &\simeq& 6.75 m_d
    \nonumber\\&&\times 
    \left( \frac{ Y_{\rm sol} }{ 10~\mbox{MPa} } \right)^{-0.71}
    \left( \frac{ \rho_{d} }{ 3.3~\mbox{g~cm}^{-3} } \right)^{0.71}
    \left( \frac{ v_{\rm imp} }{ 1~\mbox{km~s}^{-1} } \right)^{1.42}
    .\nonumber\\
\end{eqnarray}
We calculate the mass-loss rates of chondrules by 
\begin{eqnarray}
    \frac{d m_{\rm ch} }{ dx } &=& 
    m_{\rm loss} n_{\rm col}
    = m_{\rm loss} n_{d} \pi a_{\rm ch}^2 \frac{v_{\rm imp}}{v_{\rm ch}},
\end{eqnarray}
for solid chondrules.
Here, $n_{\rm col} = n_{d} \pi a_{\rm ch}^2 v_{\rm imp} / v_{\rm ch}$ is the number of collisions per unit travel length of chondrules.

We use the same model for the droplet chondrule, assuming $Y_{\rm liq}=0$.
This assumption means that the collisions between a droplet chondrule and a droplet dust particle cause the catastrophic destruction of the droplet chondrule (see Appendix \ref{sect:breakup} for the breakup criteria of collisions between droplets).
Figure \ref{fig:schematic} is a schematic summary of our collision model.
We calculate the cumulative number of the collisions during $S<1$, $N_{S}$. 
All droplet chondrules ($S = 0$) are destroyed when $N_S=1$ and we stop the simulation.
Large chondrules cause collisions during the phase transition.
During the phase transition, we remove the mass of the molten part ($(1-S) m_{\rm ch}$) from the chondrules when $N_S$ reaches one, and we reset $S$ to 1
\footnote{ 
This assumption is based on impact studies \citep[e.g.,][]{Wunnemann+2010, Bisighini+2010}.
When the thickness of the molten part is shorter than the dust size, this assumption is not good and the solid part is also ejected.
We note that this criterion for the ejection from the lower solid part would change if the molten part contained relict grains \citep[e.g.,][]{Hewins+2005,Jacquet+2021}.
This assumption does not affect our results.
}.
We neglect the effect of the impact debris \citep[][]{Jacquet&Thompson_C2014}. 
This is because we focus on the destruction of the chondrules during melting. 

\section{Results} \label{sect:results}

\subsection{Typical Evolution}\label{sect:typical}

\begin{figure*}[ht!]
    \plotone{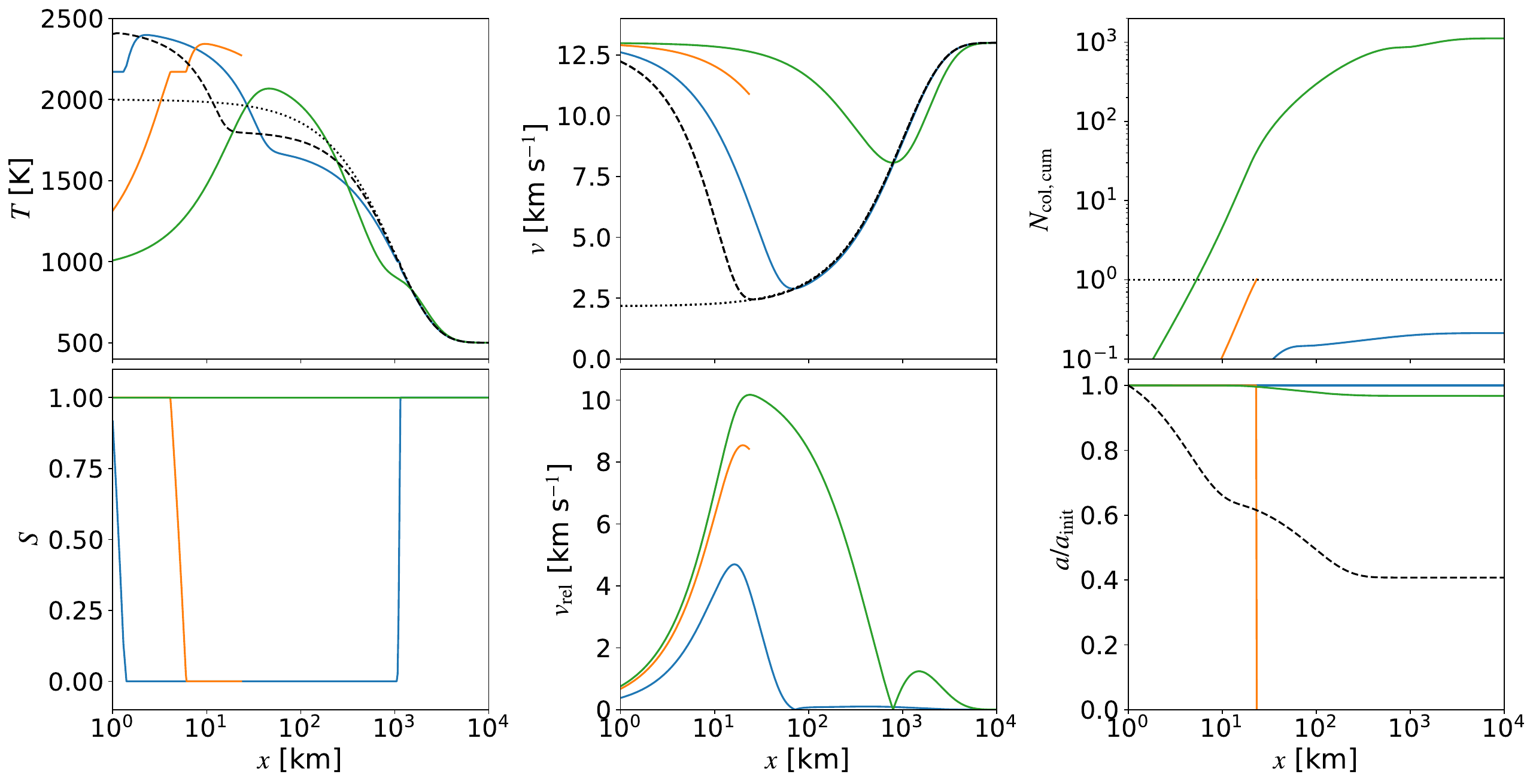}
    \caption{
        Temperature ($T$), solid mass fraction ($S$), velocity with respect to the shock front ($v$), relative velocity ($v_{\rm rel}$), cumulative number of collisions ($N_{\rm col,cum}$), and size ratio to the initial one ($a/a_{\rm init}$) are plotted against distance from the shock front in the case that $v_0=13~\mbox{km~s}^{-1}$, $\chi=0.1$, and $a_{\rm d,init}=20~\mbox{$\mu$m}$.
        The solid lines are the quantities of the chondrules. 
        The different colors indicate the different initial sizes of the chondrules, $a_{\rm ch,init}=10^{1.6} (=39.8)~\mbox{$\mu$m}$ (blue), $10^{2.2} (=158)~\mbox{$\mu$m}$ (orange), and $10^{3}~\mbox{$\mu$m}$ (green).
        In the panels of $T$ and $v$, the dotted lines are those of the gas. 
        In the panels of $T$, $v$, and $a/a_{\rm init}$, the dashed lines are those of the dust particles.
        The dotted line in the $N_{\rm col,cum}$ panel is the line for $N_{\rm col,cum}=1$.
        \label{fig:typical}}
\end{figure*}

Figure \ref{fig:typical} shows the typical evolution of chondrules and dust particles behind the shock front in the case where $v_0=13~\mbox{km~s}^{-1}$, $\chi=0.1$, and $a_{\rm d,init}=20~\mbox{$\mu$m}$.
We show the evolution of the chondrules with three different sizes.
In the case of the smallest chondrules ($a_{\rm ch,init}=10^{1.6} ~\mbox{$\mu$m}$), the chondrules heat up quickly and become droplets ($S=0$) at $x\simeq2~\mbox{km}$.
The temperature of the chondrules decreases due to cooling by radiation and evaporation after they reach the peak temperature.
The chondrules turn into supercooled droplets at $x=x_s\simeq 16~\mbox{km}$ and crystallize at $x=x_c\simeq 1.1\times10^3~\mbox{km}\simeq L$.
The velocities of the chondrules are also quickly damped, and the relative velocity between the chondrules and dust reaches a maximum at $x=16~\mbox{km}$, which is around the stopping length of the dust particles.
The maximum relative velocity is $4.7~\mbox{km~s}^{-1}$.
Then, the relative velocity becomes lower as the velocities of the chondrules are damped \citep[see ][for more details on the evolution of the relative velocity]{Matsumoto&Arakawa2023}. 
The cumulative number of the collisions is $N_{\rm col,cum}\simeq 1.8\times 10^{-2}$ at $x=x_s$ and $N_{\rm col,cum}\simeq0.20$ at $x=x_c$.
This indicates that about 80\% of the chondrules do not experience collisions during melting and most collisions between the chondrules and dust particles occur during supercooling.

After crystallization, the chondrules are not eroded by collisions.
This is because the relative velocities between the chondrules and the dust particles are very low, and there are also no collisions.
The chondrules have shorter stopping lengths than the spatial scale of the shock ($l_{\rm stop,ch}<L$), which makes relative velocities low at $x>x_c$ ($v_{\rm rel}<0.07~\mbox{km~s}^{-1}$). 
The cumulative number of the collisions is $N_{\rm col,cum}\simeq0.21$ at $x=10^4~\mbox{km}$, which is almost the same as $N_{\rm col,cum}$ at $x=x_c$.

All medium-sized chondrules ($a_{\rm ch,init}=10^{2.2} ~\mbox{$\mu$m}$) experience collisions with dust particles during melting.
These chondrules have larger cross-sections and higher relative velocities to dust particles than the smallest chondrules, which makes the cumulative number of collisions greater.
The cumulative number of collisions between chondrules and dust particles exceeds unity during melting, and all chondrules are destroyed.

The largest chondrules ($a_{\rm ch,init}=10^3~\mbox{$\mu$m}$) are not destroyed in collisions.
These chondrules do not become droplets, and they can avoid destructive collisions.
The chondrules are eroded by collisions between the solid chondrules and the dust. 
The final size and mass of the chondrules are 96.8\% and 90.6\% of the initial size and mass.
Of the total size change of the chondrules, 90\% (corresponding to $0.971a_{\rm ch}$) occurs up to $x=210~\mbox{km}$, where about 530 out of 1121 collisions occur.
The chondrules are mainly eroded by high-velocity collisions, $v_{\rm imp}>6~\mbox{km~s}^{-1}$, in the short distance. 
The effect of the low-velocity collisions in the long distance on the erosion rate is small.

We find three fates of chondrule (precursors): (1) small collisionless chondrules ($\lesssim 10^2~\mu$m) survive; (2) $10^2~\mbox{$\mu$m}\lesssim a_{\rm ch,init}\lesssim 10^3~\mbox{$\mu$m}$ chondrules are destroyed during melting; and (3) large unmelted chondrules (precursors) ($\gtrsim 10^3~\mu$m) do not melt and survive.
The survival of the large unmelted chondrules (precursors) is consistent with \cite{Nakamoto&Miura2004_LPSC}.
The survival of the small collisionless chondrule is similar to the compound chondrules formation, where the chondrule-chondrule collisions are rare \citep[][]{Ciesla2006,Arakawa&Nakamoto2019}.
The destruction of the medium-sized chondrules during melting is our key finding.

\subsection{Parameter of chondrule destruction}

\begin{figure*}[ht!]
    \plottwo{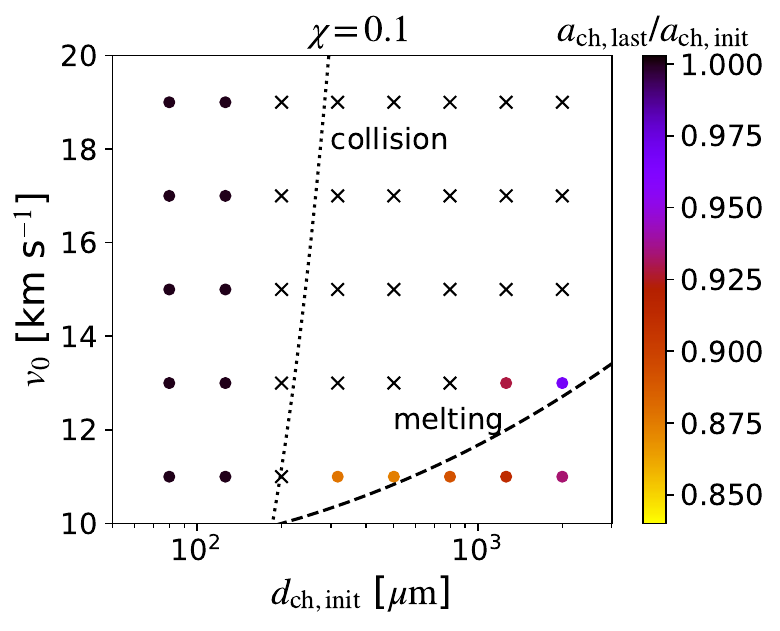}{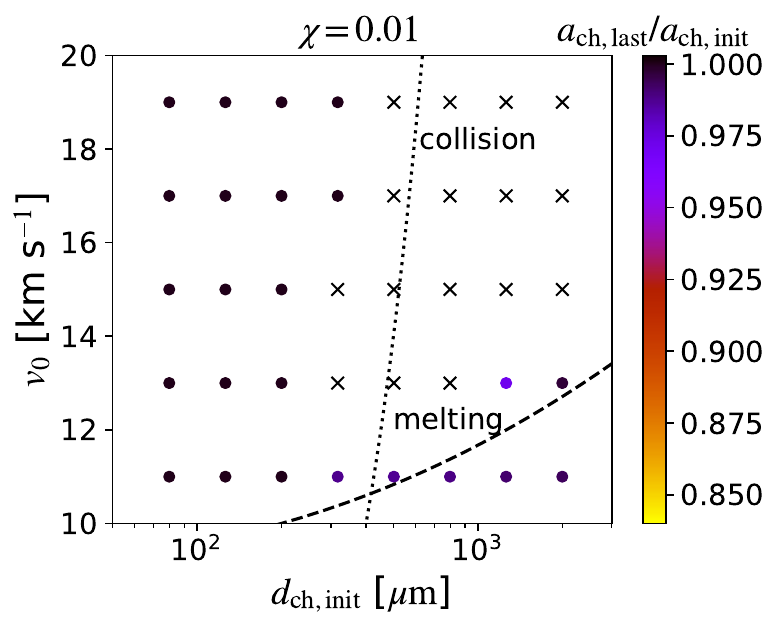}
    \caption{
        Final states of the chondrules are plotted on the initial chondrule diameter ($d_{\rm ch,init}$) and the shock velocity ($v_0$) planes.
        In the case of the chondrule survival, we plot the circle symbols, whose colors represent the size ratios of the final chondrules to the initial ones.
        The case of chondrule destruction is represented by the cross symbols.
        The dotted lines and dashed lines are the collision condition (Equations (\ref{eq:num_col_condition})) and the melting condition.
        The left and right panels show the cases of $\chi=0.1$ and $\chi=0.01$, respectively.
        \label{fig:dch_vs_ad20}}
\end{figure*}

The types of chondrule evolution are characterized by two conditions: melting and collision.
We show the final states of the chondrules on the initial chondrule diameter ($d_{\rm ch,init}=2a_{\rm ch,init}$) and the shock velocity ($v_0$) plane in Figure \ref{fig:dch_vs_ad20}.
Large chondrules do not melt at low shock velocities.
Some of small chondrules do not experience collisions. 
Chondrules that satisfy the melting and collision conditions are destroyed during melting.
The melting condition is roughly estimated by substituting the melting temperature into the equation of energy (Equation (\ref{eq:energy})).
The collision condition is estimated by integrating the number of collisions.
This integration is approximately given by $n_{\rm col} l_{\rm stop,ch}$ for small chondrules. 
Given that the number density of dust particles is the same as that in the preshock region, the collision condition is expressed as,
\begin{eqnarray}
    n_{\rm col} l_{\rm stop,ch}
    &\sim& \left( \chi \frac{m_g n_0}{m_d} \right) \pi a_{\rm ch}^2 \frac{v_{\rm imp}}{ v_{\rm ch}} 
    l_{\rm stop,ch}>1,
    \nonumber\\
    \Leftrightarrow~
    a_{\rm ch,init}&>& 
    0.9\times 10^{2}~\mbox{$\mu$m}~
    \left( \frac{v_0}{ 10\mbox{~km~s}^{-1}} \right)^{2/3}
    \nonumber\\&&\times 
    \left( \frac{\chi}{0.1} \right)^{-1/3}
    \left( \frac{a_d}{20~\mbox{$\mu$m}} \right)
    .
    \label{eq:num_col_condition}
\end{eqnarray}
This equation approximately reproduces the boundary of the collision in Figure \ref{fig:dch_vs_ad20}.

The final sizes of the unmelted chondrules are comparable to the initial sizes (the bottom right of Figure \ref{fig:dch_vs_ad20}).
The ratios of the final sizes to the initial sizes, $a_{\rm ch,last}/a_{\rm ch,init}$, are greater than 0.87.
These chondrules experience many low-velocity collisions (Section \ref{sect:typical}). 
These collisions do not cause erosion but would cause accretion \citep{Matsumoto&Arakawa2023}.

Equation (\ref{eq:num_col_condition}) shows that the typical size of the observed chondrules, $\sim 10^2$--$10^3~\mbox{$\mu$m}$ \citep[e.g.,][]{Friedrich+2015}, satisfies the collision condition.
This suggests that the observed chondrules are not formed in the shock heating events when small dust particles coexist. 
The shock events that form chondrules are different from the shock events that melt and accrete igneous rims.

\section{Conclusion and Discussion} \label{sect:conclusion}

Chondrules and igneous rims are considered to experience similar heating events in the solar nebula.
We consider the shock wave heating in their (precursors) coexistence.
We focus on the destruction of chondrules in collisions with dust particles.
We find three regimes: the survival of collisionless chondrules; the destruction of chondrules by collisions during melting; and the survival of unmolten chondrules.
The collisionless chondrules have an initial size of $\lesssim 10^2~\mbox{$\mu$m}$.
The chondrules larger than this size are destroyed by collisions if they melt, and they are not destroyed if they do not melt.
This suggests that the chondrule-forming shock events are destructive to chondrules in the presence of dust particles and that they are distinct from the rim-accreting shock events.

The key point of this work is the presence of $\sim 10~\mbox{$\mu$m}$ dust particles for the destruction of $\sim1$~mm chondrules.
If the dust particles are large and comparable to the size of the chondrules, the destruction of chondrules by collisions during melting does not occur \citep{Ciesla2006,Arakawa&Nakamoto2019}.
If the dust sizes are $\lesssim 1~\mu$m, the dust is completely evaporated and the chondrule destruction occurs if the molten chondrules experience collisions before the whole dust evaporation (Appendix \ref{sect:evap}).

\begin{acknowledgments}
    We thank Prof. A. M. Nakamura for comments on the strength of chondrules.
    Numerical simulations were in part carried out on analysis servers at Center for Computational Astrophysics, National Astronomical Observatory of Japan.
\end{acknowledgments}

\appendix
\section{Heating and cooling rates of a particle} \label{sect:model_GLL}

The heating rate via the energy transfer from the gas, $\Gamma$, is given by
\begin{eqnarray}
    \Gamma = m_g n_g | v-v_g | (T_{\rm rec} -T) C_H,
    \label{eq:Gamma}
\end{eqnarray}
where the recovery temperature, $T_{\rm rec}$, and a heat transfer function, $C_H$, are \citep[e.g.,][]{Gombosi+1986}
\begin{eqnarray}
    T_{\rm rec} &=& \frac{T_{\rm g}}{\gamma+1} 
    \left[ 2\gamma+ 2(\gamma-1) s^2 
    - \frac{\gamma-1}{ 0.5+ s^2 + (s/\sqrt{\pi}) \exp{(-s^2)} {\rm erf}^{-1}(s) } \right],
\end{eqnarray}
and
\begin{eqnarray}
    C_{\rm H} &=& \frac{\gamma+1}{\gamma-1} \frac{ k_{\rm B} }{8m_{\rm g} s^2} 
    \left[ \frac{s}{\sqrt{\pi}}\exp{(-s^2)} + \left( \frac{1}{2}+s^2\right)\erf{(s)} \right],
    \nonumber\\
\end{eqnarray}
respectively.
The radiative cooling rate is given by
\begin{eqnarray}
    \Lambda_{\rm rad} &=& \epsilon_{\rm emit} \sigma_{\rm SB} T^4 - \epsilon_{\rm emit} \sigma_{\rm SB} T_0^4,
    \label{eq:Lambda_rad}
\end{eqnarray} 
where $\epsilon_{\rm emit}$ is the emission coefficient and $\sigma_{\rm SB}=5.67\times 10^{-5}~\mbox{erg~cm$^{-2}$~K$^{-4}$~s$^{-1}$}$ is the Stefan--Boltzmann constant. 
The emission coefficient is \citep{Rizk+1991,Miura&Nakamoto2005}
\begin{eqnarray}
    \epsilon_{\rm emit} = \min{\left[ 2.1543\times 10^{-3} \left( \frac{a}{1~\mu\mbox{m}} \right)^{0.8253},1 \right]}.
    \label{eq:emission_coefficient}
\end{eqnarray}
The latent heat cooling by evaporation is the product of the evaporation rate, $J_{\rm evap}$, and the latent heat of evaporation, $L_{\rm evap}$,
\begin{eqnarray}
    \Lambda_{\rm evap} &=& J_{\rm evap} L_{\rm evap}, 
\end{eqnarray}
where $J_{\rm evap}$ is 
\begin{eqnarray}
    J_{\rm evap} &=& 691 \left( \frac{p_{\rm H_2}}{100~\mbox{dyn~cm}^{-2} } \right)^{1/2}
    \left( \frac{T }{ T_{\rm melt} } \right)^{-1/2} 
    \exp{\left( -\frac{3.17\times10^4~\mbox{K}}{T} \right)} ~\mbox{g~cm$^{-2}$~s}^{-1},
    \label{eq:J_evap}
\end{eqnarray}
and $L_{\rm evap}=1.12\times 10^{11}~\mbox{erg~g}^{-1}$ for forsterite \citep{Miura+2002}.
The ambient gas pressure, $p_{\rm H_2}$, is the summation of the thermal pressure and the ram pressure\footnote{The pressure of the net gas-grain flow on the product of the evaporation rate is unclear and not included in this study.},
\begin{eqnarray}
    p_{\rm H_2} = n_{\rm g} k_{\rm B} T_{\rm g} + \frac{1}{3} m_{\rm g} n_{\rm g} (v -v_{\rm g})^2.
\end{eqnarray}

\section{Boiling and evaporation of a dust particles} \label{sect:ad_ev}

Dust particles boil when their vapor pressure, $p_{\rm eq}$, exceeds the gas pressure, $p_{\rm H_2}$ \citep[][see also \cite{Nagahara&Ozawa1996}]{Miura+2002}, where 
\begin{eqnarray}
    p_{\rm eq} &=& 3.20 \times 10^8 \exp{\left(-\frac{6.18\times10^4~\mbox{K}}{T}\right)}~\mbox{bar}.
\end{eqnarray}
The boiling rate of a dust particle is 
\begin{eqnarray}
    -4\pi  a_{d}^2 \rho_{d} L_{\rm boil} \left(\frac{{\rm d} a_{d}}{{\rm d} x}\right)_{\rm boil} =
    \frac{4\pi a_{d}^2}{v_{d}} (\Gamma - \Lambda_{\rm rad} - \Lambda_{\rm evap} ),
    \nonumber\\
\end{eqnarray}
where $L_{\rm boil}=1.64\times 10^{11}~\mbox{erg~g}^{-1}$ is the latent heat of boiling \citep{Miura+2002}.
The size of a dust particle also decreases due to evaporation from the precursor surface, and its rate is 
\begin{eqnarray}
    -4\pi  a_{d}^2 \rho_{d} \left(\frac{{\rm d} a_{d}}{{\rm d} x}\right)_{\rm evap} = \frac{4\pi a_{d}^2}{v_{d}} J_{\rm evap}.
\end{eqnarray}

\section{breakup criteria of droplet collisions}\label{sect:breakup}

The breakup criteria of droplets in collisions are characterized by the Weber number, $\We$,
\begin{eqnarray}
    \We = \frac{2\rho a v_{\rm imp}^2 }{\sigma},
\end{eqnarray}
where $\sigma$ is the surface tension.
A droplet is destroyed when $\We$ exceeds the critical value, $\We_{\rm cr}$.
It is known that $\We_{\rm cr}\sim10$ for the droplet-flow interaction \citep[e.g.,][]{Bronshten1983,Kadono&Arakawa_M2005, Kadono+2008}.
Substituting $\We_{\rm cr}\sim10$ and $\sigma=400~\mbox{erg~cm}^{-2}$ \citep{Murase&McBirney1973}, we can get the critical velocity, 
\begin{eqnarray}
    v_{\rm cr}\sim100~\mbox{cm~s}^{-1}
    \left( \frac{ a_{\rm ch} }{ 0.1~\mbox{cm} } \right)^{-1/2}
    \left( \frac{ \We_{\rm cr} }{ 10 } \right)^{1/2}.
    \label{eq:v_cr}
\end{eqnarray}
The criterion for the collisional destruction of the droplets is different from this droplet-flow interaction criterion. 
The collision experiments of the droplets show that the outcomes of the droplets with $\We\gtrsim10$ are coalescence, bounce, and separation according to the Weber number and impact parameters \citep[e.g.,][see also \cite{Ciesla2006}]{Qian_J&Law_CK1997}.
There are some experimental and numerical studies considering collisions of droplets with $\We$ higher than $\sim10$.
\cite{Pan_KL+2009} performed experiments on head-on collisions of equal-sized droplets at high Weber numbers and showed that droplet collisions become destructive when 
\begin{eqnarray}
    \We\geq 301.0+1124.3\Oh,
    \label{eq:we_cr}
\end{eqnarray}
where the Ohnesorge number, $\Oh$, is given the ratio between the viscous and surface energies, 
\begin{eqnarray}
    \Oh = \frac{\eta}{\sqrt{2\rho \sigma a}}, 
    \label{eq:oh}
\end{eqnarray}
and $\eta$ is the viscosity.
Substituting the viscosity of the chondrules in H3 chondrites \citep[][]{Hubbard2015},
\begin{eqnarray}
    \log_{10}{\left(\frac{\eta}{1~\mbox{P}}\right)} &=& -3.55 + \frac{5084.9~\mbox{K}}{T-584.9~\mbox{K}},
    \label{eq:eta}
\end{eqnarray}
the Ohnesorge number of a chondrule becomes
\begin{eqnarray}
    \Oh = 
    6.5\times 10^{-5.55+ \frac{5084.9~{\rm K}}{T_{\rm ch}-584.9~{\rm K}} }
    \left( \frac{\rho_{\rm ch}}{ 3\mbox{~g~cm}^{-3} } \right)^{-1/2}
    \left( \frac{\sigma}{ 400\mbox{~erg~cm}^{-2} } \right)^{-1/2}
    \left( \frac{a_{\rm ch}}{ 0.1\mbox{~cm} } \right)^{-1/2}.
\end{eqnarray}
The critical Weber number begins to increase rapidly due to the contribution of the Ohnesorge number when the temperature becomes less than about 1800~K.
In this case, we can rewrite the critical velocity as,
\begin{eqnarray}
    v_{\rm cr} &\sim& 5 ~\mbox{cm~s}^{-1}
    \times 
    10^{\frac{2542.5~{\rm K}}{T_{\rm ch}-584.9~{\rm K}} }
    \left( \frac{\rho_{\rm ch}}{ 3\mbox{~g~cm}^{-3} } \right)^{-1/4}
    \left( \frac{\sigma}{ 400\mbox{~erg~cm}^{-2} } \right)^{-1/4}
    \left( \frac{a_{\rm ch}}{ 0.1\mbox{~cm} } \right)^{-3/4}.
\end{eqnarray}
We consider that the critical velocity of droplet destruction in collisions is much lower than $\sim 1~\mbox{km~s}^{-1}$, and we assume that the collisions between the droplet chondrules and droplet dust particles are destructive in the postshock region.
We note that supercooled droplets with extremely high viscosity are expected to survive high-velocity collisions \citep[][]{Arakawa&Nakamoto2019,Matsumoto&Arakawa2023}, but the survival of supercooled droplets does not significantly change the results of this paper.
This is because the critical velocity for droplet destruction becomes $1~\mbox{km~s}^{-1}$ when $T$ becomes less than $\sim 1200$~K.

\section{Effect of Dust evaporation}\label{sect:evap}

\begin{figure}[ht!]
    \plotone{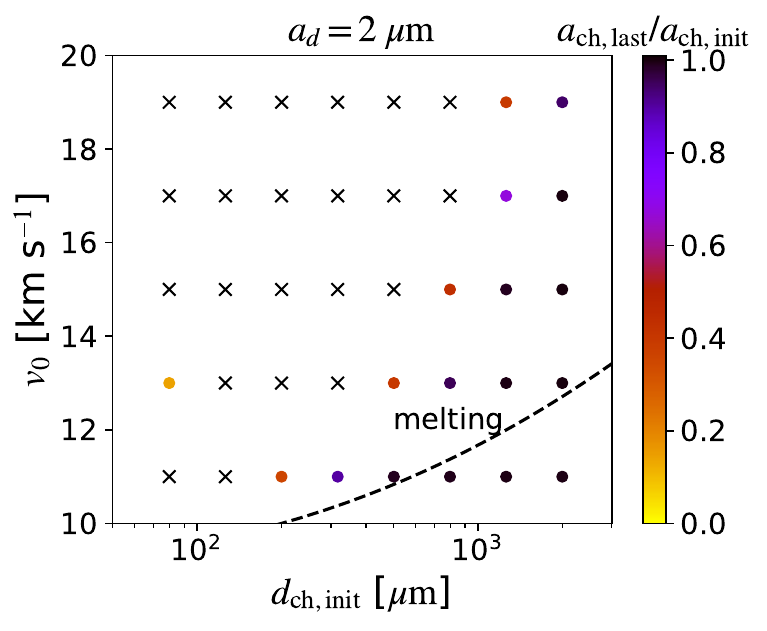}
    \caption{
        Same as Figure \ref{fig:dch_vs_ad20}, but for the case of $a_d=2~\mbox{$\mu$m}$ and $\chi=0.1$.
        The minimum size for collisional destruction is $a_{\rm ch,init}=9~\mbox{$\mu$m}$} (Equations (\ref{eq:num_col_condition})) and is satisied in all cases.
        \label{fig:dch_vs_ad2_chi1e-1}
\end{figure}

In the case of $a_d=20~\mbox{$\mu$m}$ and $n_0=10^{14}~\mbox{cm}^{-3}$, dust particles are completely evaporated when $v_0\geq17~\mbox{km~s}^{-1}$. 
When $v_0=17~\mbox{km~s}^{-1}$, the complete dust evaporation occurs at $x\simeq 2.1\times 10^2~\mbox{km}$.
In this case, molten chondrules can avoid destructive collisions when $a_{\rm ch}\gtrsim5\mbox{~mm}$.
Chondrules with $a_{\rm ch}\lesssim5\mbox{~mm}$ are destroyed by collisions. 
This is because they melt at $x<2.1\times 10^2~\mbox{km}$ and the molten chondrules collide with the dust particles before complete evaporation.

The effect of the dust evaporation can help $\sim 1$~mm chondrules to survive in the shock wave when the dust size is $\lesssim 1~\mbox{$\mu$m}$.
Figure \ref{fig:dch_vs_ad2_chi1e-1} shows the final states of the chondrules in the case of $a_d=2~\mbox{$\mu$m}$ and $\chi=0.1$, where the critical velocity for the dust evaporation is less than $10~\mbox{km~s}^{-1}$.
Since the dust particles are evaporated, there are some parameters, in which $\sim1~$mm chondrules can survive although they are molten.
This suggests that the presence of the $\lesssim 1~\mbox{$\mu$m}$ dust particles does not prevent chondrules from forming in shock waves.

\bibliography{bibtex_ym}{}
\bibliographystyle{aasjournal}

\end{document}